\documentclass[preprint2]{aastex}

\usepackage{epsfig}

\def\gx{GX~339$-$4}
\def\cx{Cyg~X$-$1}

\def\grs{GRS~1915$+$105}

\def\groquatre{GRO~J0422$+$32}

\def\gs{GS~2023$+$338}
\def\gstreize{GS~1354$-$64}
\def\1e{1E~1740.7$-$2942}

\def\xte{XTE~J1550$-$564}
\def\xteonze{XTE~J1118$+$480}

\slugcomment{Accepted for publication in ApJ Letters}

\shorttitle{Near-Infrared synchrotron emission in GX 339-4}
\shortauthors{Corbel and Fender }

\begin{document}


\title{Near-infrared synchrotron emission from the compact jet of \gx }


\author{S. Corbel}

\affil{Universit\'e Paris VII and Service d'Astrophysique, CEA Saclay, F-91191 Gif sur Yvette, France }

\email{corbel@discovery.saclay.cea.fr}

\and

\author{R.P. Fender}

\affil{Astronomical Institute `Anton Pannekoek', University of Amsterdam, 
and Center for High Energy Astrophysics, Kruislaan 403, 1098 SJ Amsterdam, 
The Netherlands }

\email{rpf@astro.uva.nl}



\begin{abstract}

We have compiled contemporaneous broadband observations of the black
hole candidate X-ray binary GX 339-4 when in the low/hard X-ray state
in 1981 and 1997. The data clearly reveal the presence of two spectral
components, with thermal and non-thermal spectra, overlapping in the
optical -- near-infrared bands. The non-thermal component lies on an
extrapolation of the radio spectrum of the source, and we interpret it
as optically thin synchrotron emission from the powerful, compact jet
in the system. Detection of this break from self-absorbed to optically
thin synchrotron emission from the jet allows us to place a firm lower
limit on the ratio of jet (synchrotron) to X-ray luminosities of $\geq
5$\%. We further note that extrapolation of the optically thin
synchrotron component from the near-infrared to higher frequencies
coincides with the observed X-ray spectrum, supporting models in which
the X-rays could originate via optically thin synchrotron emission from 
the jet (possibly instead of Comptonisation).

\end{abstract}




\section{Compact Jet from Black Hole Candidates in the low/hard X-ray State}

Radio emission from black hole X-ray binaries has long been associated
with bright radio flares ($>$ 100 mJy) starting around the onset of an
X-ray outburst. This is usually interpreted as
synchrotron emission from relativistic electrons ejected from the
system with large bulk velocities; in a few cases such jets have been
directly imaged. In the rising phase of the flares the radio
spectrum is usually flat or inverted, a characteristic of optically
thick synchrotron emission (spectral index $\alpha \ge 0$ for a flux
density S$_\nu$ $\propto$ $\nu^\alpha$), and then quickly evolves to a
decay phase characterised by optically thin emission with negative
spectral index.  For recent reviews, see Hjellming \& Han (1995),
Mirabel \& Rodr\' \i guez (1999) and Fender \& Kuulkers (2001).

By observing the persistent black hole candidates (BHCs) in our galaxy (in particular \gx\ and
\cx), it has been possible to show that they were usually associated
with weak (few mJy) radio counterparts with different properties than
the bright radio flaring transient BHCs.  Indeed, \gx\ and \cx\ are
most of the time characterized by a persistent radio source with a
flat or slightly inverted ($\alpha \ge 0$) radio spectrum
\cite{mar96,cor00}. Such radio properties have been
interpreted as arising from a conical, self absorbed compact jet (on
milli-arcsec scale), similar to those considered for flat spectrum
active galactic nuclei (AGN) (Blandford \& K$\ddot{\mathrm{o}}$nigl 1979;
Hjellming \& Johnston 1988; Falcke \& Biermann 1996).  This
interpretation has been successfully confirmed with the direct imaging
of a compact jet in \cx\ \cite{sti01}. Radio emission has also been
shown to be correlated with soft and hard X-ray emission
\cite{han98,bro99,cor00}. Observations of \gx\ have shown that
the compact jet was quenched in the high/soft state (Fender et
al. 1999; Corbel et al. 2000). Recent observations of X-ray novae have shown
that the compact jet was an ubiquitous property of BHCs in the
low/hard state (Fender 2001a; Corbel et al. 2001a and references
therein).  The compact jet seems to be quenched in the
Intermediate/Very High states, i.e. in any state where a strong soft
X-ray component exists \cite{cor01a}.

Compact jet models predict a cut-off or break to the flat or inverted
spectral component above a frequency at which the jet is no longer
self-absorbed, even at the base.  This break is observed in the
millimeter range for the compact cores of most flat spectrum AGN
(e.g. Bloom et al. 1994). To date there has been no clear detection of
such a high-frequency break in the self-absorbed synchrotron spectrum
from an X-ray binary system. Detection of such a high frequency
cut-off is important as a firm lower limit to the radiative
luminosity of the self-absorbed part of the compact jet can be
established by measuring this high frequency cut-off.  
The secondary star in \gx\ is very likely an evolved low mass
star (e.g. Shahbaz, Fender, \& Charles 2001; Chaty et al. 2002) without significant thermal
contribution in the near-infrared range. \gx\ is therefore maybe the best candidate to
look for a high frequency cut-off to the compact jet spectral
component. We, therefore, looked at all published (to our
knowledge) optical and near-infrared observations of \gx, while in the
low/hard state. We present evidence that the cut-off frequency of
the compact jet is in the near infrared range, based on observation
performed in a bright low/hard state in 1981. This is confirmed by
later observations in 1997. We then discuss the implication of this
cut-off on the energetic content of the compact jet related to the
bolometric X-ray luminosity of the system.

\section{Observations}

\subsection{A bright low/hard state of \gx\ in 1981}

Observations in 1981 revealed that \gx\ had been very active with
transitions to various X-ray states. Indeed, in March 1981, optical
observations showed that the optical counterpart was very faint (B
$\approx$ 20.34 $\pm$ 0.15 mag., V $\geq$ 19.5 mag.), and therefore
indicated that \gx\ was in the off state \cite{ilo81,ilo87}. X-ray
observations performed by the {\em Hakucho} satellite on April 7
failed to detect \gx\ with an upper limit of 15 mCrab \cite{mae84}. On
May 7$^{\mathrm{th}}$, \gx\ was already in a brighter optical state
with V $\sim$ 16.3 mag. \cite{gri81}. Subsequent optical and X-ray
observations (see below) revealed that \gx\ was in the standard
low/hard state up to June 27, where the X-ray monitoring observations
performed by {\em Hakucho} showed that \gx\ made a transition to a
high/soft state \cite{mae84}. According to Motch et al. (1983), the
peak of the low/hard state was around May 28${^\mathrm{th}}$.  

\subsubsection{Optical, near-infrared  and X-ray observations} 

When \gx\ was found to be in a bright low/hard state in May 1981,
several observations were performed in the optical and
near-infrared bands. Here we decided to concentrate on the period from
May 24$^{\mathrm{th}}$ to June 4$^{\mathrm{th}}$, as the flux from
\gx\ was observed to be similar in V and B filters, on May
28${^\mathrm{th}}$ and June 4$^{\mathrm{th}}$. Near-infrared
observations in J, H, K and L bands were performed with the ESO-3.6 m
telescope in La Silla, Chile on May 24$^{\mathrm{th}}$ by Motch,
Ilovaisky and Chevalier (1981). Using the 1.5 m Danish telescope at La
Silla, Motch et al. (1981) also performed
optical observations in the B, V, R and I bands on May
28$^{\mathrm{th}}$. Pedersen (1981) conducted U, B and V bands optical
observation on June 4$^{\mathrm{th}}$.
A conservative error of 0.1 magnitude is applied in this study to the
various optical and near-infrared measurements. An optical extinction
of A$_\mathrm{V}$ = 3.7 $\pm$ 0.3 mag., coupled with the extinction
law of Cardelli, Clayton, \& Mathis (1989), have been used to deredden
these data (Zdziarski et al. 1998).  The uncertainty on the dereddened flux estimates is
dominated by the uncertainties on the optical extinction. 
The optical and near-infrared data have been discussed in various papers, but in a
different framework than the one presented here (Motch et al. 1983; Motch et al. 1985; an 
optical extinction of A$_\mathrm{V}$ = 2 mag. was used in Motch et
al. 1985). The data used in this paper are summarized in Table~1.

An X-ray spectrum in the 1--50 keV range has been obtained by the
satellite {\em Ariel-6} on 1981, May 30-31.  The data were best fitted
with a power-law of photon-index $\sim$ 1.5 \cite{ric83}, typical for
the low/hard X-ray state.  As these X-ray data are simultaneous with
the optical and near infrared observations mentioned above, they have
been used in this paper to estimate the luminosity in soft and hard
X-rays and are plotted in Figure 2.

\subsubsection{The level of radio emission}

Hannikainen et al. (1998) first pointed out that the radio and X-ray
emission in \gx\ were strongly correlated during the low/hard state,
a correlation which has been studied in more details by Corbel et
al. (2000) using a longer extensive set of data. Using simultaneous
RXTE (PCA, HEXTE) and radio observations, we have been able to improve
significantly the quality of this study (previous works only used
BATSE and RXTE/ASM data). We observed a very strong power-law
correlation between radio, soft and hard X-ray emissions in the
low/hard state over three orders of magnitude in X-ray flux.
The correlation corresponds to the relation $F_{\rm rad} =
1.72 F_{\rm X}^{0.71}$, where
F$_{\mathrm{rad}}$ is the radio flux density (in mJy) at 8640 MHz and
F$_\mathrm{X}$ the 3--9 keV flux (in unit of 10$^{-10} \rm \, erg
\,s^{-1} \,cm^{-2}$) (Corbel et al. in prep.).  As no radio observations
were performed in 1981 (the radio counterpart was discovered
in 1994 by Sood \& Campbell-Wilson 1994), we can estimate the level of
radio emission around the time of the optical-near infrared
observations, using the X-ray spectrum obtained simultaneously by {\em
Ariel-6}. A flux of 22.3 $\times$ 10$^{-10} \rm \, erg \,s^{-1}
\,cm^{-2}$ in the 3-9 keV band is measured; the above mentionned
fitting function therefore allow us to estimate the radio flux density
of \gx\ to 16 mJy at 8640 MHz (with a conservative error of 1 mJy). As
the correlation observed in \gx\ is maintained over several years (Corbel et al. in prep.),
it is safe to assume that it represents the true level of
radio emission of \gx\ that would have been detected on 1981, May
30$^{th}$. The typical spectral index of the inverted spectral component in the
radio regime has been found to be around +0.15 during the most
sensitive radio observations performed in the low/hard state
\cite{cor00}.

\subsection{Additional observations performed during a low/hard state}

In addition to the data from 1981, we found only one other case of
near infrared observation during a low/hard state, which appear not to
be dominated by the thermal emission from the disk. It is data from
the ESO 2.2 m telescope taken on 1997, July 19$^{\mathrm{th}}$ (Chaty et al. 2002),
they are presented in Table 1. Almost simultaneously, \gx\ was
observed by the MOST radio telescope at 843 MHz at a level of 4.2~$\pm$~0.7~mJy 
on 1997, July 22$^{\mathrm{nd}}$. We also found that one
X-ray observations (discussed in Wilms et al. 1999) was made by
RXTE on 1997, July 7$^{\mathrm{th}}$. We found that the CGRO/BATSE
20-100 keV and RXTE/ASM 2-12 keV fluxes were on average a factor 1.75
fainter on July 7$^{\mathrm{th}}$ compared to July
19$^{\mathrm{th}}$. We therefore used this factor to scale the X-ray
observation at the time of the near-infrared observations as the shape
of the X-ray energy spectrum stays almost constant in the low/hard
state. Based on this RXTE spectrum and on the radio - X ray
correlation (Corbel et al. in prep.), the level of radio emission at 8640 MHz is
estimated to 5.0~$\pm$~0.1~mJy, giving a spectral index of
0.08~$\pm$~0.08 if we take into account the MOST detection.

\section{Results and discussion}

\subsection{Spectral energy distribution in the low/hard state: evidence for near-infrared synchrotron emission}

The optical and near infrared flux, corrected for interstellar
extinction, have been plotted in Figure 1 for the 1981 low/hard state
data. At first inspection it is obvious that the near infrared
datapoints clearly deviate from an extension of the optical slope. For
illustration purpose, these observations have been fitted with a
function consisting of a sum of two powerlaws, resulting in a spectral
index of --0.6 for the near-infrared range and +2.1 for the optical
data. The optical data are broadly consistent with thermal emission,
presumably from an accretion disc. On the contrary, emission in the
J, H, K and L bands clearly points to an infrared excess of
non-thermal origin.

In order to understand the nature of this infrared excess, we have
plotted in Figure 2 the spectral energy distribution from radio to
hard X-rays during the 1981 low/hard state of \gx. 
Using an extrapolation of the radio spectrum (with typical spectral index) 
up to the infrared range, we find that the L band
datapoint is compatible with a simple extension of the powerlaw
originating from the radio domain. This seems to indicate that radio
and near-infrared emission may have a common physical
origin. Considering the fact that the radio emission of \gx\ or other
BHCs in the low/hard state has been interpreted as the optically thick
synchrotron emission from a compact jet \cite{cor00,fen01a}, it is
likely that we are observing the near-infrared synchrotron emission
from the compact jet of \gx. The negative spectral index of the
near-infrared data indicates that they lie above the optically thin
break, which probably lies at a wavelength of a few microns.

In Figures 1 \& 2 we also plot the broadband data from 1997 July.  Again
the near infrared level is compatible with an extension of the flat
component originating in the radio domain. Connecting the K band
observation to the MOST detection with a powerlaw results in a
spectral index of 0.08, in agreement with the estimated spectral index
in radio range. It is interesting to note that the radio, near-infrared
and X-ray emission all varied by about the same factor ($\sim$ 4 times
fainter) between 1981 and 1997, which favors a direct link, if not a
common physical origin for all three spectral components. 
Inspection of Figure~1 may also indicate that the cut-off
frequency was at slightly higher energy in 1997, but this would
require to be verified by more observations.

In addition to \gx, there has been some indication that near infrared
and/or optical synchrotron emission from a compact outflow also took
place in few other sources. Indeed, it has been shown that the
inverted (or flat) radio spectra from \groquatre, \xteonze, \gstreize,
\xte, \grs, \gs\ (see references in Fender 2001a and Corbel et
al. 2001a) probably extend up to the near infrared - optical range,
but no high frequency break to this component has yet been directly
observed.  Such a break has been inferred at a wavelength of a few
microns for the low/hard state transient XTE~J1118+480 (Markoff, Falcke \& Fender 2001);
in this source it seems clear that, whatever the model,
there is excess flux in the near-infrared which cannot be explained by
an accretion disc alone (Hynes et al. 2000, see also Hynes et al. 2002 for XTE~J1859+226).
In \xte\ there was fairly good evidence that the break
frequency lies in the near infrared \cite{cor01a,cor01b}.
Recently, Jain et al. (2001) confirmed our
interpretation, based on their near infrared and optical monitoring of
\xte\ during the 2000 X-ray outburst.  Indeed a secondary flare
(prominent in near infrared, but also visible in optical) was
associated with the transition to the low/hard state and the
reappearance of the compact jet.  

With these observations of \gx, taken during the bright
low/hard state in 1981, we clearly detect two emission
components in the near infrared - optical domain. The optical
datapoints typically represents the thermal emission from the outer
part of the accretion disk, whereas the near infrared points
correspond to optically thin regime of the synchrotron emission from
the electron distribution. The high frequency cut-off to the flat (or
inverted) spectra from the compact jet therefore probably falls in the
near infrared range. This is the first time that such a high
frequency break is unambiguously detected in the flat (or inverted)
synchrotron spectrum from a compact jet. The observation taken during
the 1997 low/hard state also show similar cut-off (Figure 1), and
therefore favors this scenario. 

\subsection{A powerful compact jet}

The detection of a high frequency cut-off to the optically thick
spectral component is vitally important in order to estimate the total
radiative luminosity of the compact jet. 
In the rest of this paper, we only consider the optically thick
spectral component of the jet -- while optically thin synchrotron
emission may extend through the optical, UV and X-ray bands
(e.g. Markoff et al. 2001), in that energy regime disentangling its
contribution from that of thermal, Comptonised or other high-energy
components is not straightforward.  With the observed level of near
infrared emission, a spectral index of +0.15 and the high frequency
synchrotron break at a few microns ($\sim$ 10$^{14}$ Hz), the total
radiative luminosity of the compact jet of \gx\ (during the 1981
low/hard state) is about L$_\mathrm{J}$ = 10$^{35} \rm \, erg
\,s^{-1}$ (10$^{28} \rm \, W $) for a distance of 4 kpc. 
The total jet power, P$_\mathrm{J}$, can be estimated as
P$_\mathrm{J}$~=~L$_\mathrm{J}~\eta^{-1}$~F($\beta$,i), where $\eta$
represents the radiative efficiency for the jet and F($\beta$,i) a
correction factor for bulk relativistic motion (dependent on the bulk
motion velocity $\beta$ in unit of c and the angle i to the line of
sight), see discussion in Fender (2001a,b).  A value of $\eta$ = 0.05
seems a conservative estimate of the radiative efficiency of the jet,
based on minimum power requirement of the repeated ejections from
\grs\ (Fender \& Pooley 2000, Fender 2001b), which also have a flat
spectrum from radio to near-infrared. 
The effect of bulk relativistic motion
can not be precisely determined, but this results in
over-estimating the jet power only for low values of the inclination
angle (see Figure 6 in Fender 2001a).  
As the jet is both radiatively inefficient, and likely to 
have a large, possibly dominant, kinetic energy component, this represents 
a firm lower limit on the jet power.
With these limitations in mind, a likely lower limit to the total power of the compact jet in \gx\
(during the 1981 low/hard state) is estimated to be around
P$_\mathrm{J}$~=~2$\times$10$^{36} \rm \, erg \,s^{-1}$
(2$\times$10$^{29}$ W).

Based on the simultaneous {\em Ariel-6} observations (Figure~2), the
integrated X-ray luminosity of \gx\ in the 1-50 keV band is
2$\times$10$^{37} \rm \, erg \,s^{-1}$ ( 2$\times$10$^{30} \rm \, W $).  
As the X-ray luminosity of BHCs in the low/hard state is
known to peak around $\sim$ 100 keV, extrapolating the {\em Ariel-6}
X-ray spectrum up to 200 keV results in a 1-200 keV X-ray luminosity
of L$_\mathrm{X}$~=~4$\times$10$^{37} \rm \, erg \,s^{-1}$
(4$\times$10$^{30} \rm \, W $). Therefore, it is fairly safe
to say that the jet power of \gx\ is at least 5 \% of the bolometric
X-ray luminosity (which is presumed to reflect the accretion rate).
(For the observations in 1997, we obtained a total jet power  of
P$_\mathrm{J}$~=~5$\times$10$^{35} \rm \, erg \,s^{-1}$ and a 
1-200 keV X-ray luminosity L$_\mathrm{X}$~=~5$\times$10$^{36} \rm \, erg \,s^{-1}$,
resulting in a fraction of 8 \%).
In \xte, Corbel et al. (2001a) derived similar value for the jet to
accretion luminosities ratio. Based
on similar arguments, Fender (2001a) obtained similar value for the
compact jet of \cx\ using an extrapolation of the radio-mm spectrum up
to the near infrared range. A higher ratio ($\sim$ 20\%) has been
found for the recently discovered X-ray transient \xteonze\ (Fender et
al. 2001).  The general trend which is starting to emerge from these
repeated multi-wavelength observations of compact jet sources is that
the power in the jet may be a significant fraction (similar in all
sources ?) of the total accretion luminosity.

The base of the jet is probably located very close to, or co-existent
with, the corona (Fender et al. 1999), which would explain why there
is a very strong coupling between the hard X-ray emission from the
corona and the radio emission from the compact jet
\cite{han98,cor00}.  This may arise via Comptonisation of
electrons in the base of the jet or, more radically, it may be that
the X-ray emission observed in the low/hard spectral state is also
optically thin synchrotron emission directly from the jet (Markoff et
al. 2001). In Figure 2, as well as the radio -- infrared -- optical data,
we also plot the contemporaneous X-ray data. Extrapolation of the
optically thin synchrotron component in the near-infrared agrees
remarkably well with the measured X-ray spectra. Despite the limited
data sets, we consider this to be a support to the model of
Markoff et al. (2001), but this interpretation will need to be confirmed 
by further broadband observations and will be discussed in another paper (Markoff et al. in prep.).

\section{Conclusions}

The compilation of data presented here for the black hole candidate
X-ray binary GX 339-4 clearly reveals the presence of an additional,
apparently non-thermal, spectral component in the optical /
near-infrared spectral bands, above the thermal emission expected from
the accretion flow/disc. We interpret this non-thermal spectral
component as optically thin synchrotron emission from just above the
frequency at which the self-absorbed spectrum observed in the radio
band breaks to optically thin (ie. it is no longer self-absorbed in
any part of the jet). This is the strongest evidence to date that this
break lies in the spectral region of the near-infrared to optical
bands. We further note that extrapolation of this non-thermal
component is coincident with the X-ray spectrum, supporting models in
which the X-rays  could arise via  optically thin synchrotron emission from the jet
(instead of Comptonisation).

However, the interpretation of the X-ray spectra is necessarily
complex, and the relative contributions of disc, Comptonised,
synchrotron and other components is not straightforward to separate.
Our approximate measurement of the high-frequency extent of the
self-absorbed synchrotron component does however give us a very firm
and conservative lower limit to the power associated with the
jet. Comparing this with the measured X-ray emission, we find that the
jet requires {\em at least} 5\% of the total accretion power. This is
further evidence for the ubiquity of powerful jets in the low/hard
spectral state of accreting black holes.

\acknowledgements

The authors acknowledge useful discussions with Sera Markoff.


\clearpage



\figcaption[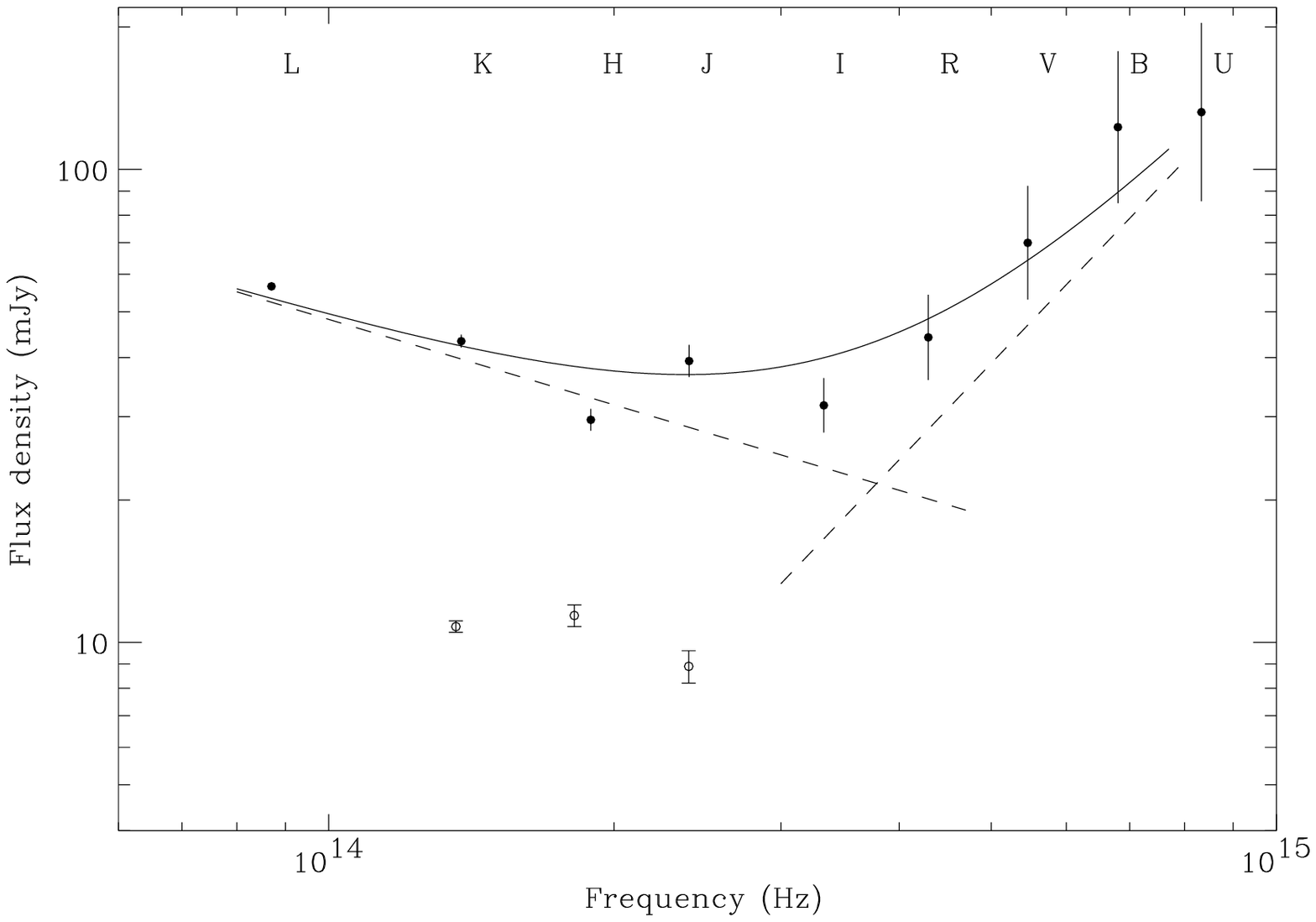]{Optical and near infrared magnitudes dereddened with an optical extinction 
of A$_\mathrm{V}$ = 3.7 $\pm$ 0.3 mag. Filled circles are observations for the 1981 low/hard state and open
circles for the 1997 low/hard state. For 1981, the data have been fitted with the sum (continuum line) 
of two powerlaw components (dotted lines) with a spectral index of --0.6   for the near-infrared range 
and + 2.1 for the optical range.\label{fig1}}

\figcaption[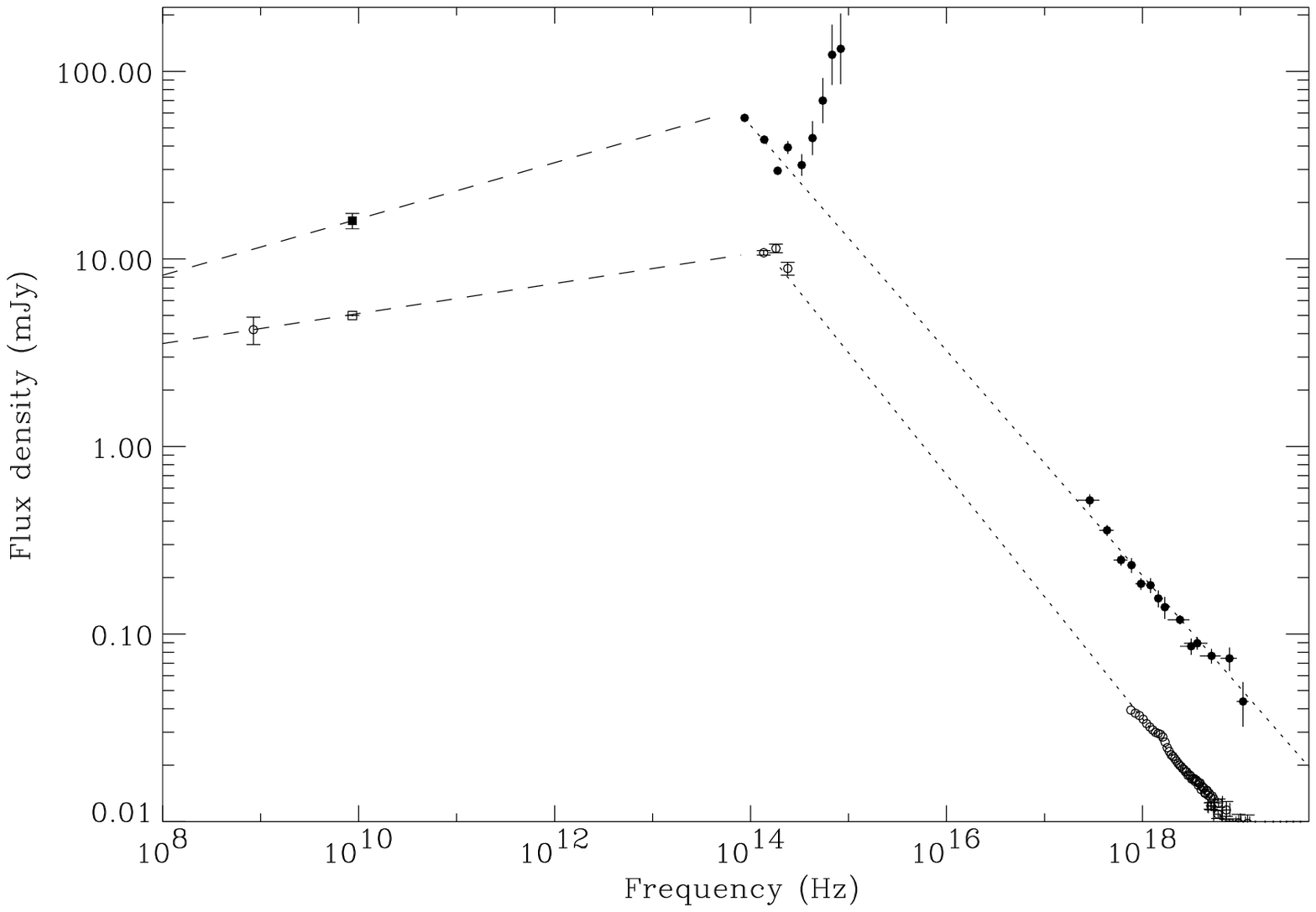]{Broadband radio-infrared-optical and X-ray spectra of GX~339--4 during a
low/hard state in 1981 (filled symbols) and 1997 (open symbols). Filled and open circles represent
the various measurements obtained from quasi-simultaneous observations. The square symbols is the 
level of radio emission at 8640 MHz estimated from the measured X-ray fluxes based on the existing 
correlation between radio and X-ray emission (Corbel et al. 2002). The long-dashed and short-dashed 
lines correspond to self absorbed and optically thin regimes in the jet respectively (spectral indexes of +0.15  and --0.6 for 1981; +0.08 and --0.65 for 1997). These two lines are for illustrative purpose only 
as a smooth transition is expected between the two regimes (e.g. Markoff et al. 2001). \label{fig2}}





\clearpage

\begin{deluxetable}{cllcrc}
\footnotesize
\tablecaption{Log of optical and near-infrared observations. \label{tbl-1}}
\tablewidth{0pt}
\tablehead{
\colhead{Band} & \colhead{Date}   & \colhead{Telescope}   &
\colhead{Flux (mag.)} & \colhead{Flux (mJy) \tablenotemark{a}} & \colhead{References}
}
\startdata

U & 1981, June 4 & Danish 1.5m & 16.2 & 132.1  $\pm$ 59.1 & Pedersen (1981)   \\
B & 1981, June 4 & Danish 1.5m & 16.3 & 122.8  $\pm$ 46.4 & Pedersen (1981) \\
V & 1981, June 4 & Danish 1.5m & 15.5 &  69.9  $\pm$ 19.6 & Pedersen (1981)  \\
R & 1981, May 28 & Danish 1.5m & 14.8 &  44.1  $\pm$ 9.2  & Motch et al. (1981) \\
I & 1981, May 28 & Danish 1.5m & 13.9 &  31.7  $\pm$ 4.2  & Motch et al. (1981)  \\
J & 1981, May 24 & ESO 3.6m    & 12.6 &  39.3 $\pm$  3.1  & Motch et al. (1981) \\
H & 1981, May 24 & ESO 3.6m    & 12.1 &  29.6 $\pm$  1.6  & Motch et al. (1981) \\
K & 1981, May 24 & ESO 3.6m    & 10.9 &  43.3 $\pm$  1.4 & Motch et al. (1981) \\
L & 1981, May 24 & ESO 3.6m    &  9.4 &  56.6 $\pm$  0.9 & Motch et al. (1981) \\
\hline
J & 1997, July 19 & ESO 2.2m   & 14.2 &  8.9  $\pm$  0.7 & Chaty et al. (2002) \\
H & 1997, July 19 & ESO 2.2m   & 13.1 &  11.4 $\pm$  0.6 & Chaty et al. (2002) \\
K & 1997, July 19 & ESO 2.2m   & 12.4 &  10.8 $\pm$  0.3  & Chaty et al. (2002) \\
\enddata
\tablenotetext{a}{Dereddened fluxes using an optical extinction of
A$_\mathrm{V}$ = 3.7 $\pm$ 0.3 mag., a conservative error of 0.1 mag.
is assumed.}

\end{deluxetable}

\clearpage

\begin{figure}
\plotone{Corbel_GX339_Fig1.ps}
\label{fig1}
\end{figure}

\clearpage

\begin{figure}
\plotone{Corbel_GX339_Fig2.ps}
\label{fig2}
\end{figure}

\end{document}